\documentclass[conference]{IEEEtran}
\IEEEoverridecommandlockouts
\usepackage{cite}
\usepackage{amsmath,amssymb,amsfonts}
\usepackage{algorithmic}
\usepackage{graphicx}
\usepackage{textcomp}
\usepackage{xcolor}
\usepackage{multirow}
\usepackage[super]{nth}
\usepackage{multirow}
\usepackage{verbatim}
\usepackage{threeparttable}
\def\BibTeX{{\rm B\kern-.05em{\sc i\kern-.025em b}\kern-.08em
    T\kern-.1667em\lower.7ex\hbox{E}\kern-.125emX}}
\begin{document}

\makeatletter
\newcommand{\linebreakand}{%
  \end{@IEEEauthorhalign}
  \hfill\mbox{}\par
  \mbox{}\hfill\begin{@IEEEauthorhalign}
}
\makeatother
 
\title{Relationship Between Mood, Sleepiness, and EEG Functional Connectivity by 40 Hz Monaural Beats
\footnote{{\thanks{This work was supported by the Institute of Information \& Communications Technology Planning \& Evaluation (IITP) grant, funded by the Korea government (MSIT) (No. 2019-0-00079, Artificial Intelligence Graduate School Program (Korea University)) and was partly supported by the Institute of Information \& communications Technology Planning \& Evaluation (IITP) grant, funded by the Korea government (MSIT) (No. 2021-0-02068, Artificial Intelligence Innovation Hub).}
}}
}

\author{\IEEEauthorblockN{Ha-Na Jo}
\IEEEauthorblockA{\textit{Dept. of Artificial Intelligence} \\
\textit{Korea University} \\
Seoul, Republic of Korea \\ 
hn\_jo@korea.ac.kr} \\

\and

\IEEEauthorblockN{Young-Seok Kweon}
\IEEEauthorblockA{\textit{Dept. of Brain and Cognitive Engineering} \\
\textit{Korea University} \\
Seoul, Republic of Korea \\
youngseokkweon@korea.ac.kr} \\

\linebreakand 

\IEEEauthorblockN{Gi-Hwan Shin}
\IEEEauthorblockA{\textit{Dept. of Brain and Cognitive Engineering} \\
\textit{Korea University} \\
Seoul, Republic of Korea \\
gh\_shin@korea.ac.kr}

\and

\IEEEauthorblockN{Heon-Gyu Kwak}
\IEEEauthorblockA{\textit{Dept. of Artificial Intelligence} \\
\textit{Korea University} \\
Seoul, Republic of Korea \\
hg\_kwak@korea.ac.kr}

\and

\IEEEauthorblockN{Seong-Whan Lee}
\IEEEauthorblockA{\textit{Dept. of Artificial Intelligence} \\
\textit{Korea University} \\
Seoul, Republic of Korea \\
sw.lee@korea.ac.kr}

}

\maketitle

\begin{abstract}
Monaural beats are known that it can modulate brain and personal states. However, which changes in brain waves are related to changes in state is still unclear. Therefore, we aimed to investigate the effects of monaural beats and find the relationship between them. Ten participants took part in five separate random sessions, which included a baseline session and four sessions with monaural beats stimulation: one audible session and three inaudible sessions. An electroencephalogram (EEG) was recorded and participants completed pre- and post-stimulation questionnaires assessing mood and sleepiness. As a result, an audible session led to positive mood and arousal compared to other conditions. From the neurophysiological analysis, statistical differences in frontal-central, central-central, and central-parietal connectivity were observed only in the audible session at 40 Hz. Furthermore, a significant correlation was identified between sleepiness and EEG power in the temporal and occipital regions. These results suggested a more detailed relation for stimulation to change its personal state. These findings have implications for applications in areas such as cognitive enhancement, mood regulation, and sleep management.
\end{abstract}

\begin{small}
\textbf{\textit{Keywords--electroencephalogram, mood, sleepiness, functional connectivity, monaural beats;}}\\
\end{small}

\section{INTRODUCTION}
Research activities in the fields of cognitive neuroscience and psychology have evolved. At first, the primary focus was on interpreting bio-signal patterns in trying to understand their complex state \cite{ref2, ref6}. The main purpose was to apply this understanding to the classification and interpretation of cognitive and emotional states \cite{clf, ref3}. Recently, a flow of studies focused on purposefully leading to EEG signals for communicative purposes \cite{thetaFCduringtask2}. Today, there are researchers devoted to actively using EEG patterns to modify user states in addition to merely inducing them. Our study aims to investigate how external auditory stimuli can influence EEG connectivity and subjective states in listeners, contributing to our understanding of the relationship between auditory stimuli, EEG patterns, and individual experiences.

\begin{table*}[h]
\centering
\caption{Difference between pre- and post-questionnaire of five stimulation groups}
\label{table1}
\resizebox{\textwidth}{!}{%
\tiny
\begin{tabular}{lcccccccccc}
\hline
\multicolumn{1}{c}{\multirow{2}{*}{\textbf{}}} &
  \multicolumn{2}{c}{\textbf{NS}} &
  \multicolumn{2}{c}{\textbf{AB}} &
  \multicolumn{2}{c}{\textbf{IB-f}} &
  \multicolumn{2}{c}{\textbf{IB-d}} &
  \multicolumn{2}{c}{\textbf{IB}} \\ \cline{2-11} 
\multicolumn{1}{c}{} & Mean  & SD   & Mean  & SD   & Mean  & SD   & Mean  & SD   & Mean  & SD   \\ \hline
\textit{SSS}         & 0.50  & 0.67 & -0.50 & 1.09 & -0.30 & 1.10 & 0.80  & 0.87 & 0.20  & 0.40 \\
\textit{BRUMS}       &       &      &       &      &       &      &       &      &       &      \\
\; Anger                & 0.40  & 0.66 & -0.20 & 1.16 & 0.30  & 1.26 & 0.50  & 1.36 & 0.50  & 1.91 \\
\; Tension              & 0.40  & 1.28 & 0.50  & 1.18 & -0.10 & 0.83 & -0.20 & 1.53 & -0.20 & 0.87 \\
\; Depression           & -0.10 & 1.13 & -0.40 & 1.85 & -0.30 & 0.45 & 0.20  & 0.60 & 0.20  & 0.97 \\
\; Vigor                & -0.90 & 1.64 & -0.10 & 2.34 & -1.00 & 2.48 & -0.50 & 1.11 & -0.30 & 0.78 \\
\; Fatigue              & 0.50  & 1.62 & -0.10 & 2.58 & 0.80  & 0.97 & 1.20  & 2.22 & 0.30  & 2.00 \\
\; Confusion            & -0.10 & 0.83 & -0.30 & 0.64 & 0.20  & 1.40 & 0.40  & 1.01 & -0.60 & 1.28 \\
\; Happy                & -1.10 & 1.64 & -0.70 & 2.32 & -0.30 & 0.78 & -0.40 & 0.91 & -0.20 & 1.88 \\
\; Calmness             & -0.10 & 1.04 & 0.30  & 2.53 & -0.10 & 1.57 & -0.00 & 1.89 & -0.40 & 1.28 \\ \hline
\end{tabular}}
\begin{tablenotes}
\item SSS=Stanford Sleepiness Scale, BRUMS=Brunel Mood Scale, NS=no stimulation, AB=audible monaural beats, IB-f=inaudible monaural beats in frequency, IB-p=inaudible monaural beats in power, IB=inaudible monaural beats in frequency and power, and SD=standard deviation.
\end{tablenotes}
\end{table*}

Our research focuses on monaural beats, widely used auditory stimuli for specific EEG patterns. These beats combine two different frequencies into one audio signal, effectively modulating neural activity \cite{introBB}. They have a well-documented history in cognitive neuroscience, utilized in relaxation, meditation, cognitive enhancement, and mood regulation studies, revealing EEG entrainment patterns and their effects on subjective experiences.

Recent research in this field has limitations, including a focus on a single set of auditory conditions, neglecting the inaudible frequency range \cite{ref5}. Studies also often primarily analyze EEG patterns and overlook potential secondary effects on emotions and sleepiness \cite{hana}, leaving the link between EEG activity and emotional states largely unexplored.

Our study addresses these limitations comprehensively. We administered two distinct subjective questionnaires and correlated them with EEG patterns using the functional connectivity approach. In previous studies, there were cases where stimuli were presented during task performance to assess cognitive abilities \cite{thetaFCduringtask2}. However, in our research, we conducted separate segments to ensure that task performance and stimulus-induced EEG activity did not interfere with each other. We also considered the inaudible frequency range, enabling us to explore the diverse impacts of monaural beats under various conditions. By connecting auditory stimuli, EEG responses, and their effects on subjective experiences, we contribute to an understanding of auditory interventions for state modulation.

\section{METHODS} 

\subsection{Experimental Procedure}
Included in this study were ten participants, comprising five males and five females, with a mean age of 26.1 years (standard deviation $\pm$ 3.4 years). None of the participants had a medical background indicating claustrophobia or hearing impairment. Before commencing the experiment, ethical approval was obtained from the Institutional Review Board at Korea University (KUIRB-2022-0222-01), and written informed consent was obtained from each participant.

The participants were prepared for electroencephalogram (EEG) recording in a soundproof booth \cite{ref7, ref8}. Five sessions involving different stimulation conditions were administered in a random sequence. Each session comprised five minutes for completing two questionnaires (Stanford Sleepiness Scale (SSS) and Brunel Mood Scale (BRUMS)) before and after the stimulation, followed by ten and two minutes of rest \cite{sss, brums}. Participants kept their eyes closed throughout the experiments. Afterward, the simulation session, EEG recording, and pre-processing were set in the same way as the previous study \cite{hana}.

\subsection{Neurophysiological Analysis}
To analyze the spatial information of EEG data, we segmented the brain into five distinct regions: frontal, central, temporal, parietal, and occipital regions. Subsequently, we applied the following analysis to each of these regions.

\subsubsection{Power spectral density (PSD)}
The utilization of PSD is instrumental because it allows us to effectively examine and quantify the frequency composition and power distribution within a given signal, which is essential for understanding the underlying dynamics of various physiological and neural processes \cite{ref4}.

\subsubsection{Phase lag index (PLI)}
PLI is a quantification of the asymmetry of the phase difference distribution between neural signals around non-zero phase delay synchronization, which is used in this study to investigate directional functional connectivity in the brain, and provides more accurate results by avoiding zero-phase delay interactions.

\subsection{Statistical Analysis}

\subsubsection{Paired \textit{t}-test}
A paired \textit{t}-test is a statistical method used to determine if there is a significant difference between two related groups, such as the same group tested at different times or under different conditions. We employed this method to assess the statistical significance of differences between each group. The significance level was set at \textit{p}$<$0.05.

\subsubsection{Pearson correlation}
The Pearson correlation coefficient, often denoted as Pearson's $r$, is a statistical measure that quantifies the linear relationship between two continuous variables. It assesses the strength and direction (positive or negative) of the association between these variables. The significance level was set at \textit{p}$<$0.05.

\begin{figure*}[thpb]
  \centering
  \includegraphics[width=\textwidth]{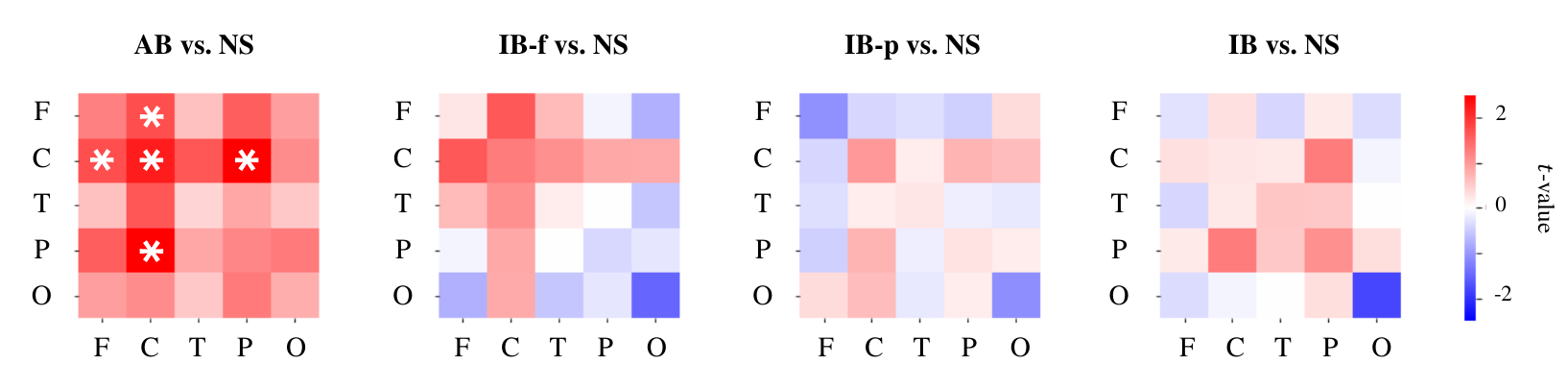}
  \caption{Differences in functional connectivity between regions of interest compared to no stimulation session (NS). The brain region is divided into five group: F=frontal, C=central, T=temporal, P=parietal, and O=occipital. AB=audible monaural beats, IB-f=inaudible monaural beats in frequency, IB-p=inaudible monaural beats in power, and IB=inaudible monaural beats in frequency and power. * represents the statistical significance (\textit{p}$<$0.05). }
  \label{figure1}
\end{figure*}

\section{RESULTS}

\subsection{Questionnaire}
In our study, we calculated the differences in questionnaire responses obtained before and after the stimulus session. Subsequently, we conducted the statistical analyses for each group, as shown in table \ref{table1}. 

In the SSS, the AB group showed significantly different values compared to the NS group (\textit{p}$=$0.03). However, no significant differences were observed in comparisons between other groups (\textit{p}$>$0.05).

Additionally, we analyzed the BRUMS, which consists of a total of 9 factors. In the `happy' factor, the AB group showed significantly larger values compared to the NS group ($p<$0.009). While there were differences in the magnitudes of responses among other groups, these differences did not reach statistical significance (\textit{p}$>$0.05).

\begin{figure}[t]
  \centering
  \includegraphics[width=240pt]{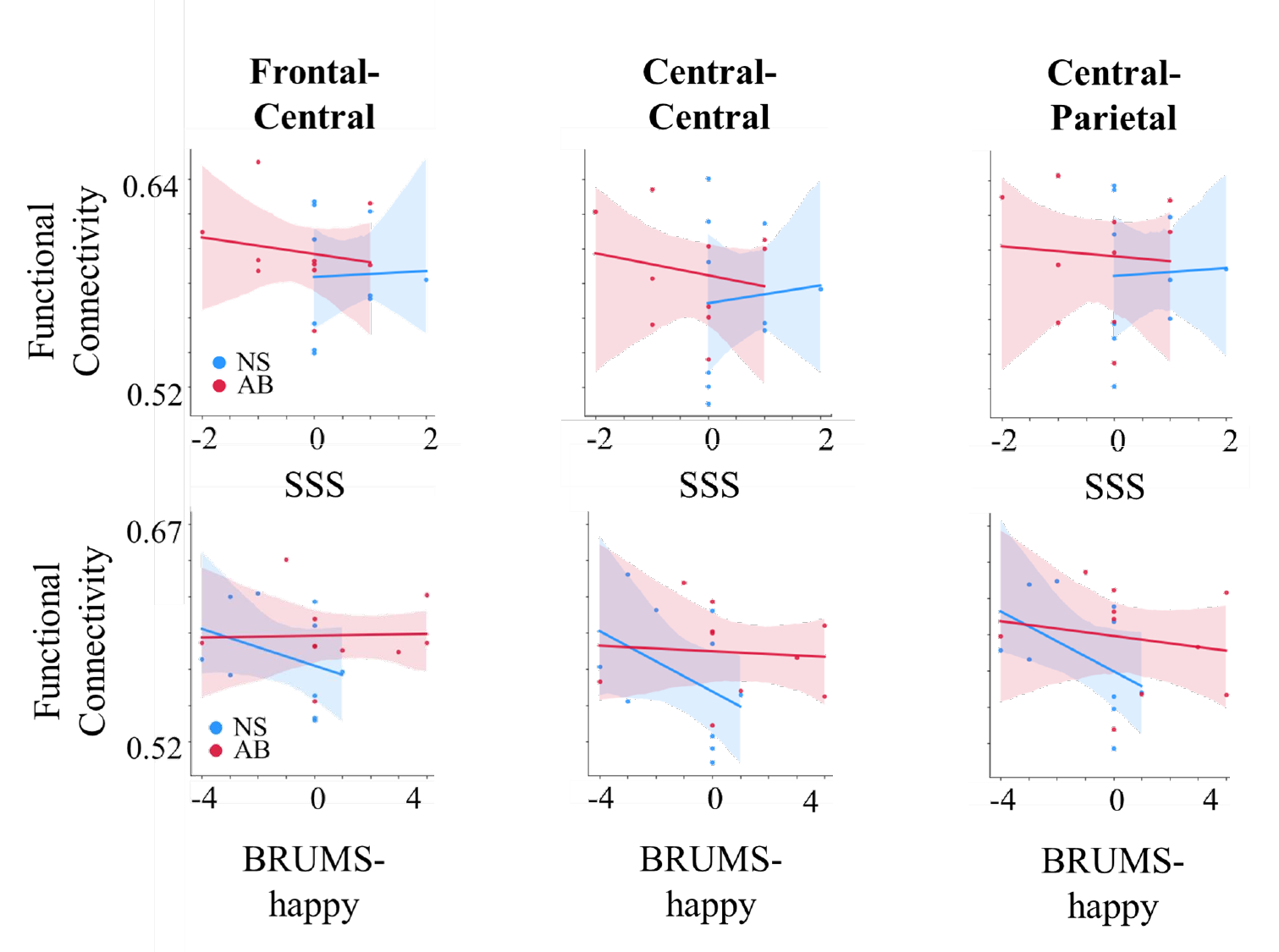}
  \caption{Correlation between questionnaire scores differences (post - pre) and EEG functional connectivity. NS=no stimulation, AB=audible monaural beats, SSS=Stanford Sleepiness Scale, and BRUMS=Brunel Mood Scale.}
  \label{figure2}
\end{figure}

\subsection{EEG Functional Connectivity}
We calculated channel-specific functional connectivity for each group and then aggregated these values within predefined five regions. Subsequently, we compared the AB, IB-f, IB-p, and IB groups with the NS group and conducted statistical analyses, as shown in Fig. \ref{figure1}. Our results revealed that in the comparison between AB and NS, there were significantly higher levels of connectivity in frontal-central (\textit{t}$=$1.73, \textit{p}$=$0.05), central-central (\textit{t}$=$2.21, \textit{p}$=$0.02), and central-parietal (\textit{t}$=$2.53, \textit{p}$=$0.01) regions. While other regions also showed positive connectivity differences in AB compared to NS, these differences did not reach statistical significance (\textit{p}$>$0.05). Similarly, for IB-f, IB-p, and IB groups, there was a trend towards negative connectivity difference compared to NS, but these differences also were not statistically significant (\textit{p}$>$0.05).

\subsection{Correlation Between Questionnaire and Connectivity}
Furthermore, we conducted an analysis focusing on the AB group, where significant results were observed, to explore the relationship between subjective states as indicated by questionnaires and EEG patterns. Firstly, We calculated the drowsiness and mood levels difference by subtracting pre- from post-questionnaire values, where negative values mean reduced drowsiness. Secondly, we explored correlation coefficients between SSS differences and functional connectivity, as well as between the BRUMS `happy' factor differences and functional connectivity, as shown in Fig. \ref{figure2}. However, in all cases, no statistically significant differences were found (\textit{p}$>$0.05).

Additionally, we also compared the questionnaires with PSD, again concentrating on the AB group where prior studies revealed significant results \cite{hana}. The results showed a negative correlation with statistical significance in the AB group between SSS-temporal power (\textit{r}$=$-0.44, \textit{p}$=$0.05) and SSS-occipital power (\textit{r}$=$-0.75, \textit{p}$=$0.01), as shown in Fig. \ref{figure3}. In contrast, other combinations of the AB group and values from the NS group were statistically insignificant (\textit{p}$>$0.05).

\begin{figure}[t]
  \centering
  \includegraphics[width=240pt]{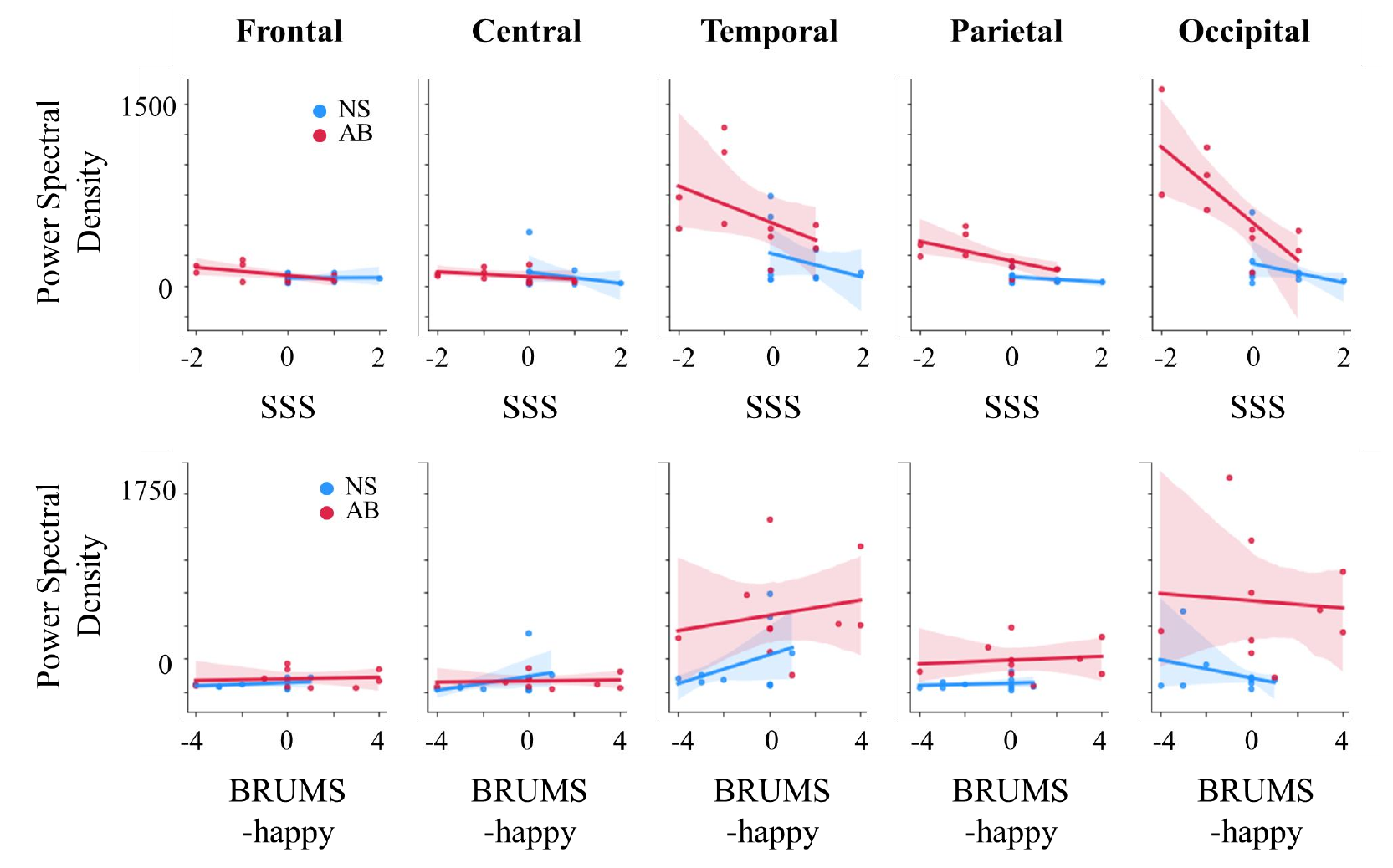}
  \caption{Correlation between questionnaire scores differences (post - pre) and EEG power spectral density. NS=no stimulation, AB=audible monaural beats, SSS=Stanford Sleepiness Scale, and BRUMS=Brunel Mood Scale.}
  \label{figure3}
\end{figure}

\section{DISCUSSION}
In our experiment, we explored the influence of auditory stimuli on both brain connectivity and subjective experiences. We categorized groups based on audible and inaudible conditions and had them fill out two self-report questionnaires. Our results revealed that only the AB group exhibited robust EEG functional connectivity at a 40 Hz, while the audible condition was associated with improved positive mood and arousal.

Our results confirm that the 40 Hz auditory stimulation in the AB condition can induce changes in subjective states. Specifically, it increases feelings of happiness and prevents drowsiness. This outcome contrasts with prior studies that involved listening to 40 Hz monaural beats for 20 to 30 minutes, where memory enhancement occurred but negative emotions significantly increased \cite{mood40hz}. Notably, the frequency used in our study to induce stability differs from the delta or theta bands employed in previous research \cite{moodothers, ref9}. However, in contrast to prior studies, our research employed an optimal 10-minute stimulation period, allowing for the maintenance of negative emotions while inducing positive moods \cite{10min}. Furthermore, to the best of our knowledge, there is limited existing research on the relationship between 40 Hz auditory beats and drowsiness. Our study unveils a novel finding, indicating that in the AB condition, 40 Hz monaural beats can effectively prevent drowsiness.

From a neurophysiological perspective, our study unveiled that the 40 Hz AB condition can create stronger EEG connectivity associated with the central regions at the targeted frequency. One previous study examined EEG power and connectivity in the theta frequency band, where they observed an increase in power but no changes in connectivity \cite{thetaFCbasic}. Another study with the theta band compared subjects listening to auditory stimuli during tasks to a control group and found significant connectivity increases in three regions \cite{thetaFCduringtask}. However, task-related EEG measurements complicate attributing observed connectivity changes solely to stimulation. Importantly, our study shows significant EEG connectivity while controlling for all variables except auditory stimulation.

Lastly, we discovered a significant correlation between the reduction in drowsiness levels post-auditory stimulation and EEG activity in the temporal and occipital regions. Our results suggest that the drowsiness reduces as EEG power is increased by monaural beats. This interpretation aligns with the known phenomenon that drowsiness or sleep is associated with an increase in lower frequencies and a decrease in higher frequencies \cite{inertia}. It was also in line with the invasive experiment results that the arousal state has a bigger gamma power compared to drowsiness states in all brain regions \cite{gammainsleep}. We found similar results in two regions (temporal and occipital) with short-time non-invasive EEG.

\section{CONCLUSION}
We conducted an investigation into the impact of audible 40 Hz monaural beat stimulation, limiting our focus to auditory conditions. Our findings indicate that under these conditions, this stimulation significantly reduces drowsiness, induces positive emotions, and enhances connectivity near central brain regions. These findings can integrate with artificial intelligence and facilitate the creation of personalized auditory stimulation protocols, optimizing drowsiness reduction, positive emotion induction, and central brain region connectivity enhancement, thus offering new possibilities for the modulation of subjective states and advancing neurophysiological research \cite{ref1, ref10}.

\bibliographystyle{IEEEtran}
\bibliography{reference}

\end{document}